\documentclass[aps,prb,english,floatfix,reprint,twocolumn,showpacs,superscriptaddress,longbibliography,10pt,tightenlines]{revtex4-1}

\usepackage[utf8]{inputenc}
\usepackage{amssymb,amsmath}
\usepackage{graphicx}
\usepackage{natbib}
\usepackage{multirow}
\usepackage{braket}
\usepackage{subcaption}
\usepackage{ragged2e}
\DeclareCaptionJustification{justified}{\justifying}
\usepackage[final]{hyperref} 
\hypersetup{
	colorlinks=true,       
	linkcolor=red,        
	citecolor=red,        
	filecolor=magenta,     
	urlcolor=blue         
}

\usepackage[dvipsnames]{xcolor}

\begin{document}

\title{False vacuum decay in the 1+1 dimensional $\varphi^4$ theory}

\author{D. Sz{\'a}sz-Schagrin}
\affiliation{Department of Theoretical Physics, Institute of Physics,\\ Budapest University of Technology and Economics, H-1111 Budapest, M{\H u}egyetem rkp. 3}
\affiliation{BME-MTA Momentum Statistical Field Theory Research Group, Institute of Physics,\\ Budapest University of Technology and Economics, H-1111 Budapest, M{\H u}egyetem rkp. 3}
\author{G. Tak{\'a}cs}
\affiliation{Department of Theoretical Physics, Institute of Physics,\\ Budapest University of Technology and Economics, H-1111 Budapest, M{\H u}egyetem rkp. 3}
\affiliation{BME-MTA Momentum Statistical Field Theory Research Group, Institute of Physics,\\ Budapest University of Technology and Economics, H-1111 Budapest, M{\H u}egyetem rkp. 3}
\affiliation{MTA-BME Quantum Correlations Group (ELKH), Institute of Physics,\\ Budapest University of Technology and Economics, H-1111 Budapest, M{\H u}egyetem rkp. 3}
\date{12th June 2022}

\begin{abstract}
The false vacuum is a metastable state that can occur in quantum field theory, and its decay was first studied semi-classically by Coleman. In this work we consider the 1+1 dimensional $\varphi^4$ theory, which is the simplest model that provides a realisation of this problem. We realise the decay as a quantum quench and study the subsequent evolution using a truncated Hamiltonian approach. In the thin wall limit, the decay rate can be described in terms of the mass of the kink interpolating between the vacua in the degenerate limit, and the energy density difference between the false and true vacuum once the degeneracy is lifted by a symmetry breaking field, a.k.a. the latent heat. We demonstrate that the numerical simulations agree well with the theoretical prediction for several values of the coupling in a range of the value of the latent heat, apart from a normalisation factor which only depends on the interaction strength.  
\end{abstract}

\maketitle

\section{Introduction}

Tunnelling in quantum field theory a.k.a.~the decay of the false vacuum was first investigated using a semi-classical approach in the groundbreaking work by Coleman\cite{Coleman1977, Coleman1977-2}. Starting with the quantum field stuck in a metastable state called the \textit{false vacuum}, bubbles of the true ground state of the theory (\textit{true vacuum}) nucleate via quantum tunnelling. The nucleated bubbles subsequently expand driven by the energy difference between the true and false vacuum and the released energy (a.k.a.~latent heat) also results in a sea of particle excitations filling the newly formed domains of the true vacuum. In the original work by Coleman\cite{Coleman1977, Coleman1977-2}, the rate of bubble nucleation was computed in a semi-classical approximation to the path integral using instantons. Recently it gathered additional attention due to indications of metastability of the Higgs vacuum in the Standard Model \cite{Elias-Miro2012}. Furthermore, recent advances in experimental technology (e.g. in trapped ultra-cold atoms) bring direct laboratory study of the phenomenon within reach.\cite{Billam2019,Billam2020,LunNg2020, Billam2021,Abel2021} 

More generally, the advances in experiments in the past few decades have promoted the out-of-equilibrium dynamics of quantum many-body systems to the forefront of research in condensed matter physics.\cite{Hofferberth_2007,Trotzky_2012,Gring_2012,Cheneau_2012,Meinhert_2013,Langen_2013,Fukuhara_2013,Kaufman2016} A paradigmatic and experimentally realisable protocol for non-equilibrium dynamics is the so-called quantum quench \cite{Calabrese2006,Calabrese2007}, where the system is initially prepared in equilibrium such as a thermal state or a ground state of some Hamiltonian. At the initial time $t = 0$ one or more parameters of the theory are suddenly changed, driving the system out of equilibrium subject to subsequent unitary time evolution. The decay of the false vacuum can be naturally implemented as a quantum quench by preparing the system in the false vacuum state as the initial state and studying the resulting time evolution. In the condensed matter context, recently the phenomenon was also studied in quantum spin chains\cite{Rutkevich1999, Lagnese2021, Pomponio2022}.

Non-equilibrium time evolution of non-trivially interacting quantum field theories is rather nontrivial to describe, requiring the use of suitable approximations, both analytic and numerical. For the 1+1 dimensional $\varphi^4$ model, which is the textbook example of a simple interacting quantum field theory, one class of methods is semi-classical approximations such as the mean-field approach\cite{Sotiriadis_2010}, or the truncated Wigner approximation\cite{Polkovnikov2003,POLKOVNIKOV2010}, both of which are limited to the regime of sufficiently weak interactions. An alternative method is provided by the truncated Hamiltonian approach (THA) which can be used for stronger interactions\cite{Szasz-Schagrin2022}. 

THA was first invented to study relevant perturbations of minimal conformal field theories\cite{Yurov_1989,1991IJMPA...6.4557Y,1991CoPhC..66...71L}, later extended to perturbations of other conformal field theories \cite{Feverati_1998,Konik_2015}, and also to perturbations of the free massive fermion \cite{Fonseca_2001}. Truncated Hamiltonian methods suitable for $\varphi^4$ were developed in the works\cite{Coser_2014,2015PhRvD..91b5005H,Rychkov_2015,Rychkov_2016,Bajnok_2016}, including also for higher space-time dimensions\cite{2015PhRvD..91b5005H}. An alternative approach to Hamiltonian truncation is provided by light-cone conformal truncation \cite{2016JHEP...07..140K,2020arXiv200513544A}. Truncated Hamiltonian methods proved efficient in simulating the full quantum out-of-equilibrium dynamics in 1+1 dimensional quantum field theories.\cite{Rakovszky_2016,2017PhLB..771..539H,2018ScPP....5...27H,2018PhRvL.121k0402K,2019JHEP...08..047H,2019PhRvA.100a3613H,2021PhRvD.104b1702K,Szasz-Schagrin2022}; for the case of perturbed conformal theories, an efficient algorithm including a MATLAB implementation was made publicly available recently.\cite{2022CoPhC.27708376H} 

In this paper we apply the THA built upon a massive free boson basis\cite{Rychkov_2015, Rychkov_2016, Bajnok_2016} to study quantum quenches involving decay of the false vacuum the 1+1 dimensional $\varphi^4$ theory, using the implementation developed in our previous work\cite{Szasz-Schagrin2022}, and determine the tunnelling rate per unit volume which can be compared directly to theoretical predictions\cite{Voloshin1985}.

The outline of the paper is as follows. In Section \ref{sec:theory} we give a brief overview of the theory of the decay of the false vacuum. Section \ref{sec:2dphi4} introduces the formulation of false vacuum decay as a quantum quench, while Section \ref{sec:THA} specifies implementation of the non-equilibrium time evolution using the truncated Hamiltonian approach. The detailed results of our investigations are presented in Section \ref{sec:results}, while Section \ref{sec:conclusions} contains our conclusions. 

\section{Decay of the false vacuum}\label{sec:theory}

Here we briefly review the aspects of the decay of the false vacuum necessary for our investigations, including features specific for $1+1$ space-time dimensions. Following Coleman\cite{Coleman1977, Coleman1977-2}, false vacuum decays via bubble nucleation initiated by quantum fluctuations. The decay is dominated by spherically symmetric bubbles. Due to the finite energy of the walls (surface tension), bubbles smaller than a critical size only appear as short-lived quantum fluctuations. However, bubbles larger than a critical radius can form as stable field configurations which then expand driven by the surplus vacuum energy density in the false vacuum compared to the true vacuum. In his seminal work\cite{Coleman1977} Coleman considered tunnelling in a scalar field theory with a potential $U(\varphi)$ which has a global minimum corresponding to a stable ground state, and a metastable local minimum corresponding to the false vacuum. In the semi-classical approximation barrier penetration is described in terms of the instanton bounce $\varphi_{\text{I}}$ which is a spherically symmetric solution to the Euclidean equation of motion:
\begin{equation}
    \left(\frac{\partial^2}{\partial \tau^2} + \nabla^2 \right)\varphi = \frac{\partial U}{\partial\varphi}
\end{equation}
satisfying appropriate boundary conditions. The tunnelling rate per unit volume $V$ is then given by the formula
\begin{equation}\label{eq:coleman}
    \gamma=\frac{\Gamma}{V} = A \exp{\left[-\frac{1}{\hbar}S_{\text{E}}\right]}
\end{equation}
where $S_{\text{E}}$ is the Euclidean action of the instanton bounce, and the amplitude $A$ can be expressed with the determinant of quantum fluctuations in the instanton background (note that it requires a careful treatment of zero modes). 

The calculation simplifies considerably if the thickness of the walls is much smaller than the radius of the critical bubble, which is called the thin wall limit. Writing the scalar potential as 
\begin{equation}
    U(\varphi)=U_0(\varphi)+\varepsilon \Delta U(\varphi)
\end{equation}
where the term $U_0$ has two degenerate minima corresponding to vacua of equal energy density. The vacuum degeneracy is explicitly broken by switching on $\varepsilon>0$ and the thin wall limit corresponds to the limit of small $\varepsilon$.\cite{Coleman1977} In 1+1 dimensions, the bubbles in the thin wall limit take the form of a kink-antikink pair delimiting a region with the true vacuum in its interior, and the diameter of the critical bubble is determined by simple energy conservation:
\begin{equation}\label{eq:resonant-bubble}
    a_* = \frac{2M}{\mathcal{E}}\,,
\end{equation}
where $M$ is the kink mass and $\mathcal{E}$ is the energy density difference between the false and true vacuum (latent heat):
\begin{equation}
    \mathcal{E} = \frac{1}{L}(E_{\text{FV}}-E_{\text{TV}})
\end{equation} 
In the thin wall limit, the action of a bubble with diameter $a$ is
\begin{equation}
    S(a)=\pi a M-\frac{\pi a^2}{4}\mathcal{E}
\end{equation}
which has its stationary point for $a=a_*$. As a result, nucleated bubbles are dominantly of the size $a_*$, and the instanton action determining the tunnelling rate is
\begin{equation}
    S_{\text{E}} = \frac{\pi M^2}{\mathcal{E}}
\end{equation}
It is possible to go beyond the semi-classical limit to include quantum corrections \cite{Coleman1977-2}. Moreover, in 1+1 dimensions these were evaluated exactly in the thin wall limit by Voloshin\cite{Voloshin1985} with the result
\begin{equation}\label{eq:voloshin}
    \gamma = \frac{\mathcal{E}}{2\pi} \exp{\left[-\frac{\pi M^2}{\mathcal{E}}\right]}\,,
\end{equation}
where $M$ is the exact renormalised kink mass and $\mathcal{E}$ is the exact quantum energy density difference between the false and true vacua. Similar results were obtained for tunnelling in the quantum Ising spin chain\cite{Rutkevich1999}, and were recently verified by numerical simulation of the spin chain dynamics.\cite{Lagnese2021}

\section{Vacuum decay as a quantum quench in the 1+1-dimensional $\varphi^4$ theory}\label{sec:2dphi4}

The action of $\varphi^4$ theory in the symmetry broken phase is given by
\begin{align}
S[\varphi]=\int d^2x \Big[ &\frac{1}{2} \left(\partial_t\varphi\right)^2-\frac{1}{2} \left(\partial_x\varphi\right)^2
\nonumber\\
&+\frac{m^2}{2} \varphi^2
-\frac{g}{6} \varphi^4+\varepsilon\varphi\Big]    
\end{align}
which corresponds to setting
\begin{equation}
    U_0(\varphi)=-\frac{m^2}{2}\varphi^2+\frac{g}{6}\varphi^4\qquad \Delta U(\varphi)=-\varphi\,.
\end{equation}
In the absence of explicit symmetry breaking (i.e., $\varepsilon=0$)the classical ground states are
\begin{equation}
\varphi_\pm=\pm\sqrt{\frac{3 m^2}{2g}}
\end{equation}
with kink/antikink excitation of mass \begin{equation}
    M=\frac{\sqrt{2}m^3}{g}
\end{equation} interpolating between them. Switching on a nonzero $\varepsilon$ makes $\varphi_+$ the true ground state, while $\varphi_-$ becomes the false vacuum (note that there positions are also slightly shifted).

At the quantum level the two vacuum states and their characteristic parameters such as the value of the order parameter and the vacuum energy density splitting acquire quantum corrections, and the same is true for the kink mass $M$.

We investigate the false vacuum decay by setting up a quantum quench protocol. Denoting the quantum Hamiltonian of the $\varepsilon=0$ theory with $H$, we initialise the system in the ground state $|\Psi_-\rangle$  of $H$ which becomes the false vacuum for $\varepsilon>0$: 
\begin{align}
&\ket{\Psi(0)} = \ket{\Psi_{-}}\nonumber\\
&H \ket{\Psi_{-}} = E_0 \ket{\Psi_{-}} \quad ,\quad \braket{\Psi_{-}|\varphi|\Psi_{-}}<0.
\end{align}
At the initial time $t = 0$ we switch on $\varepsilon>0$ and so the post-quench Hamiltonian is
\begin{equation}
    H_\varepsilon = H -\varepsilon \int dx\; \hat\varphi(x)\,,
\end{equation}
resulting in the unitary time evolution for $t>0$:
\begin{equation}
    \ket{\Psi(t)} = e^{-i H_\varepsilon t}\ket{\Psi(0)}
\end{equation}
The time evolution of an observable $\hat{\mathcal{O}}$ is given by the expectation value
\begin{equation}
    \braket{\hat{\mathcal{O}}(t)} := \braket{\Psi(t)|\hat{\mathcal{O}}|\Psi(t)}
\end{equation}
We consider the time evolution of the order parameter i.e. the expectation value of the field $\hat\varphi$, which we parameterise via the combination
\begin{equation}\label{eq:F}
    F(t) = \frac{\braket{\hat\varphi(t)}+\braket{\hat\varphi(0)}}{2\braket{\hat\varphi(0)}}\,,
\end{equation}
inspired by the study of vacuum decay in the spin chain setting by Lagnese et al.\cite{Lagnese2021}. Neglecting corrections of the vacuum expectation values of the field from the presence of the $\varepsilon$, the decay of the false vacuum corresponds to the change of $F(t)$ from $1$ to $0$, making it a convenient quantity to monitor the progression of the decay process.

\section{Applying the truncated Hamiltonian approach to vacuum decay}\label{sec:THA}

Now we turn to the application of truncated Hamiltonian approach to the time evolution starting from a false vacuum. The Hamiltonian can then be written as
\begin{equation}\label{eq:hamilton}
H_\varepsilon = H_{\text{KG}}^m + \int dx :\left(- m^2 \hat{\varphi}^2 + \frac{g}{6}\hat{\varphi}^4 -\varepsilon \hat{\varphi}\right):_m,
\end{equation}
where 
\begin{equation}\label{eq:HKGm}
H_{\text{KG}}=\int dx :\left(\frac{1}{2}\hat{\pi}^2+\frac{1}{2}(\partial_x\hat{\varphi})^2+\frac{m^2}{2}\hat{\varphi}^2\right):_m
\end{equation}
is the Klein-Gordon Hamiltonian of mass $m$, and the fields satisfy the equal time commutation relations
\begin{equation}
    \left[\hat{\pi}(t,x), \hat{\varphi}(t,x') \right] = -i \delta(x-x')\,,
\end{equation}
while $:\dots:_m$ denotes normal ordering with respect to the modes of the free Klein-Gordon field of mass $m$.

To implement the truncated Hamiltonian approach (THA), we consider the system in a finite volume $L$ with periodic (or anti-periodic) boundary condition. Working in units $m = 1$, the system can be characterised by the following dimensionless parameters:
\begin{align}
l &= m L\quad,\quad \bar g = \frac{g}{m^2}\quad\text{and}\quad
    \bar\varepsilon = \frac{\varepsilon}{m^2}\,.
\end{align}
The finite volume Hamiltonian then takes the form
\begin{align}
H_{\text{KG}}(L)&=\int\limits_0^L dx :\left(\frac{1}{2}\hat{\pi}^2+\frac{1}{2}(\partial_x\hat{\varphi})^2+\frac{m^2}{2}\hat{\varphi}^2\right):_{m,L} + E_0(L) 
\nonumber\\
H_\varepsilon(L)&= H_{\text{KG}}^m + g_0(l) V_0 + g_2(l) V_2 + g_4(l) V_4 - \varepsilon V_1
\nonumber\\
V_n &= \int_{0}^{l}dx~ :\hat{\varphi}^{n}:_{m,L}
\end{align}
Here $:\dots:_{m,L}$ denotes normal ordering with respect to the free bosonic modes with mass $m$ in a finite volume $L$, 
\begin{equation}
    E_0(l) = \int_{-\infty}^{\infty}\frac{d\theta}{2\pi}\cosh\theta\log\left(1-e^{l\cosh\theta}\right)\,,
\end{equation}
and the finite volume couplings $g_i(l)$ are related to infinite volume couplings as\cite{Bajnok_2016}
\begin{align}
g_0(l) &= -m^2z^\pm(l) -m^2\frac{3\log 2}{8\pi} + \frac{g}{2}\tilde{z}^\pm(l)^2
\nonumber\\
g_2(l) &= g\tilde{z}^\pm(l) - m^2 \qquad g_4 = \frac{g}{6} &
\label{couplings}
\end{align}
with
\begin{align}
z^+(l) &= \int_0^\infty\frac{d\theta}{\pi}\frac{1}{e^{l\cosh\theta} - 1}\qquad z^-(l) = 2z^+(2l) - z^+(l)
\nonumber\\
\tilde{z}^\pm(l) &= z^\pm(l) + \frac{\log2}{4\pi}
\end{align}

where the $\pm$ superscript refers to (anti)periodic boundary conditions $\hat{\varphi}(x + L) = \pm\hat{\varphi}(x)$.

The next step is to separate the zero mode using the "minisuperspace" method \cite{Rychkov_2016,Bajnok_2016}, to take into account the main effect of symmetry breaking (including the tunnelling) which highly improves the convergence of the method. Writing  
\begin{equation}
    \hat{\varphi}(x) = \hat{\varphi}_0 + \tilde{\varphi}(x)\,,\quad \hat{\varphi}_0=\frac{1}{L}\int_0^L dx\hat{\varphi}(x)  
\end{equation}
where $\hat{\varphi}_0$ is the zero mode, while $\tilde{\varphi}(x)$ the non-zero mode of the field, and separating the Hilbert space accordingly as 
\begin{equation}
\mathcal{H} = \mathcal{H}^\text{mini} \otimes \tilde{\mathcal{H}}
\end{equation}
the Hamiltonian can be decomposed as
\begin{align}
H_{\varepsilon} = &\tilde H_{\text{KG}}^m + H_{\varepsilon}^{\text{mini}} + \int_0^L dx~\Big[g_0 + g_2:\tilde\varphi(x)^2: 
\nonumber\\ 
& + g_4\left(:\tilde\varphi(x)^4: + 6:\tilde\varphi(x)^2:\hat{\varphi}_0^2 + 4:\tilde\varphi(x)^3:\hat{\varphi}_0\right)\Big]
\nonumber\\
H_{\varepsilon}^{\text{mini}} = & L \Bigg[\frac{1}{2}:\hat{\pi}_0^2: + \frac{m^2}{2}:\hat{\varphi}_0^2: + g_2:\hat{\varphi}_0^2: + g_4:\hat{\varphi}_0^4: 
\nonumber\\ 
& \quad - \varepsilon:\hat{\varphi}_0: \Bigg]\,,
\end{align}
where $\hat{\pi}_0$ is the zero mode conjugate momentum, and $\tilde H_{\text{KG}}^m$ denotes the free Klein-Gordon Hamiltonian with the zero mode omitted. We note that this step is only necessary for periodic boundary conditions, since there is no zero mode in the anti-periodic case.

Representing the Hamiltonian on the Fock space of the Klein-Gordon model with mass $m$ in finite volume $L$ with (anti)periodic boundary conditions, its matrix elements can be explicitly evaluated. The space decomposes in sectors of different total momentum; in our subsequent computations, we only need the sector with zero total momentum. 

The space is made finite dimensional by introducing an ultraviolet cut-off separately in the minisuperspace and the space of non-zero modes. In the minisuperspace $\mathcal{H}^\text{mini}$ the procedure is to diagonalise numerically the zero mode Hamiltonian and keep a suitably large number of the lowest lying eigenstates; we chose a cut-off for this numerical diagonalisation such that the energy levels kept after this diagonalisation can be considered numerically exact. For the non-zero modes, we impose an upper energy cut-off $\Lambda$ in the total energy computed with the KG part of the Hamiltonian. The cut-off $\Lambda$ is parameterised as 
\begin{equation}
    \frac{\Lambda}{m} = \frac{4 \pi n_{\text{max}}}{l}
    \label{eq:ncutoff}
\end{equation}
where $n_{\text{max}}$ is a dimensionless parameter which can be interpreted as to the maximum momentum quantum number which is allowed to be filled when neglecting $m$.

The spectrum of the Hamiltonian obtained in this way still depends on the UV cut-off. The leading order dependence on the non-zero mode cut-off $\Lambda$ can be eliminated by a renormalisation of the Hamiltonian:\cite{Rychkov_2016} 
\begin{align}\label{eq:effham}
    H_{\varepsilon}^{\text{RG}} = &H_{\varepsilon} + \int_0^L dx~\Big[\kappa_0 + \kappa_2:\tilde\varphi(x)^2: \nonumber\\ 
& + \kappa_4\left(:\tilde\varphi(x)^4: + 6:\tilde\varphi(x)^2:\varphi_0^2 + 4:\tilde\varphi(x)^3:\varphi_0\right)\Big]
\end{align}
where explicit expressions for the $\kappa_n$ are given in the literature\cite{Rychkov_2016,Szasz-Schagrin2022}. The zero-mode part, however, can be considered essentially exact; the only condition is to keep sufficient number of states to be consistent with the non-zero mode cut-off $\Lambda$.

\section{Results}\label{sec:results}
To evaluate \eqref{eq:voloshin} it is necessary to know the values of the kink mass and of the latent heat.
\subsection{Kink mass}
\begin{figure}[h!]
    \centering
    \includegraphics[width = \linewidth]{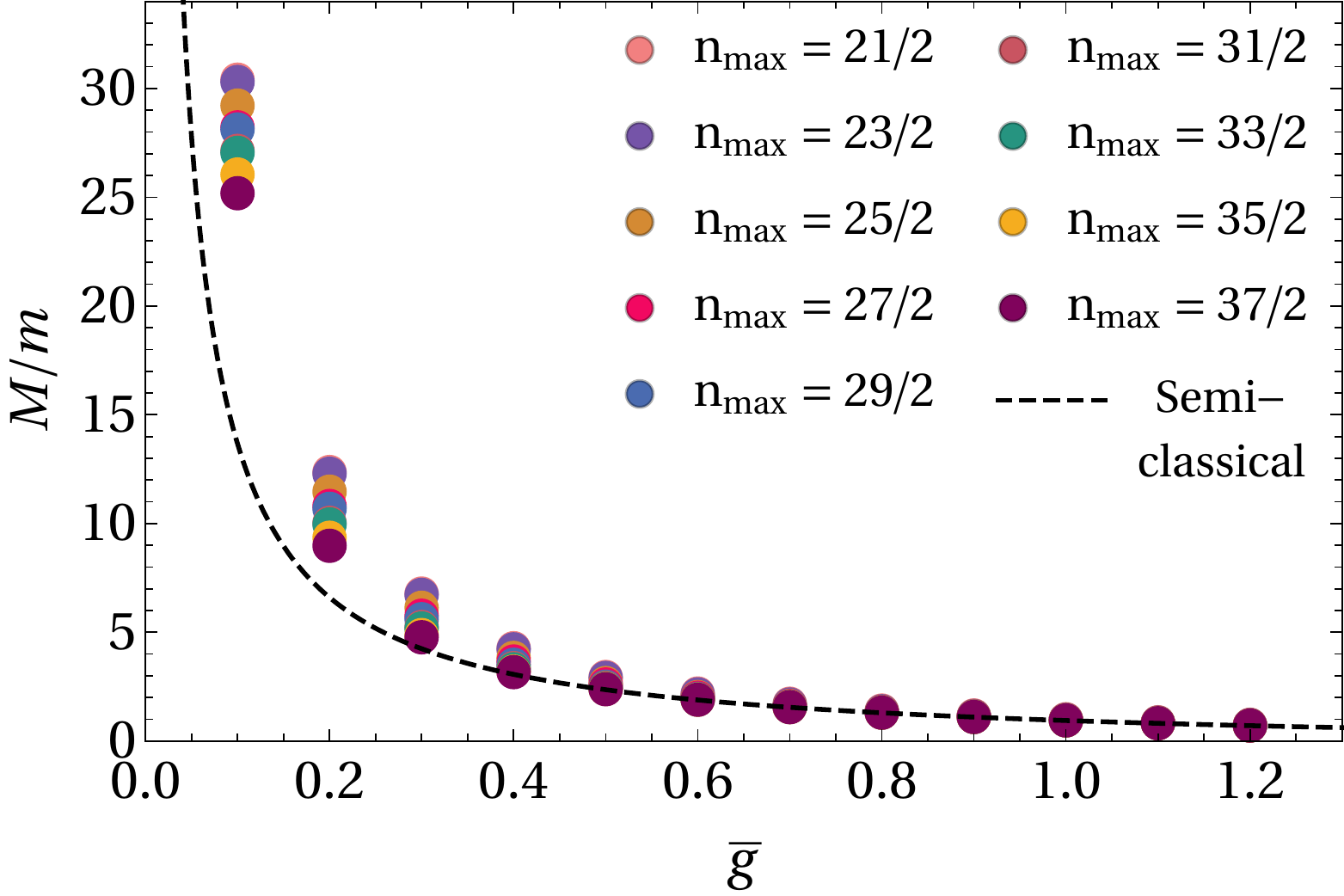}
    \caption{The kink masses as obtained via THA (markers) together with the semi-classical prediction. The different colours denote different values of the cut-off $n_{\text{max}} = 21/2, ... 37/2$ corresponding to cc. 1000-100000 basis states.}
    \label{fig:kinkmass}
\end{figure}

The kink mass can be computed by setting $\bar\varepsilon=0$, and computing the difference between the lowest levels in the anti-periodic and periodic sectors, which correspond to a stationary kink and the vacuum state, respectively. Note that for anti-periodic boundary conditions there are no zero modes, so the procedure outlined above simplifies. Fig. \ref{fig:kinkmass} shows the $g$-dependence of the kink mass\cite{Bajnok_2016} for various values of the cut-off $n_{\text{max}}$ together with the semi-classical prediction\cite{dhn2}
\begin{equation}
    M = \frac{\sqrt{2}m^3}{g} - \sqrt{2}m\left(\frac{3}{2\pi}-\frac{1}{4\sqrt{3}}\right) + O(g)
\end{equation}
Note that for small couplings the kink mass is large and therefore it has a strong dependence on the UV cut-off, while for larger couplings the kink masses agree very well with the semi-classical prediction.

\subsection{Latent heat}
\begin{figure}
    \centering
    \includegraphics[width = \linewidth]{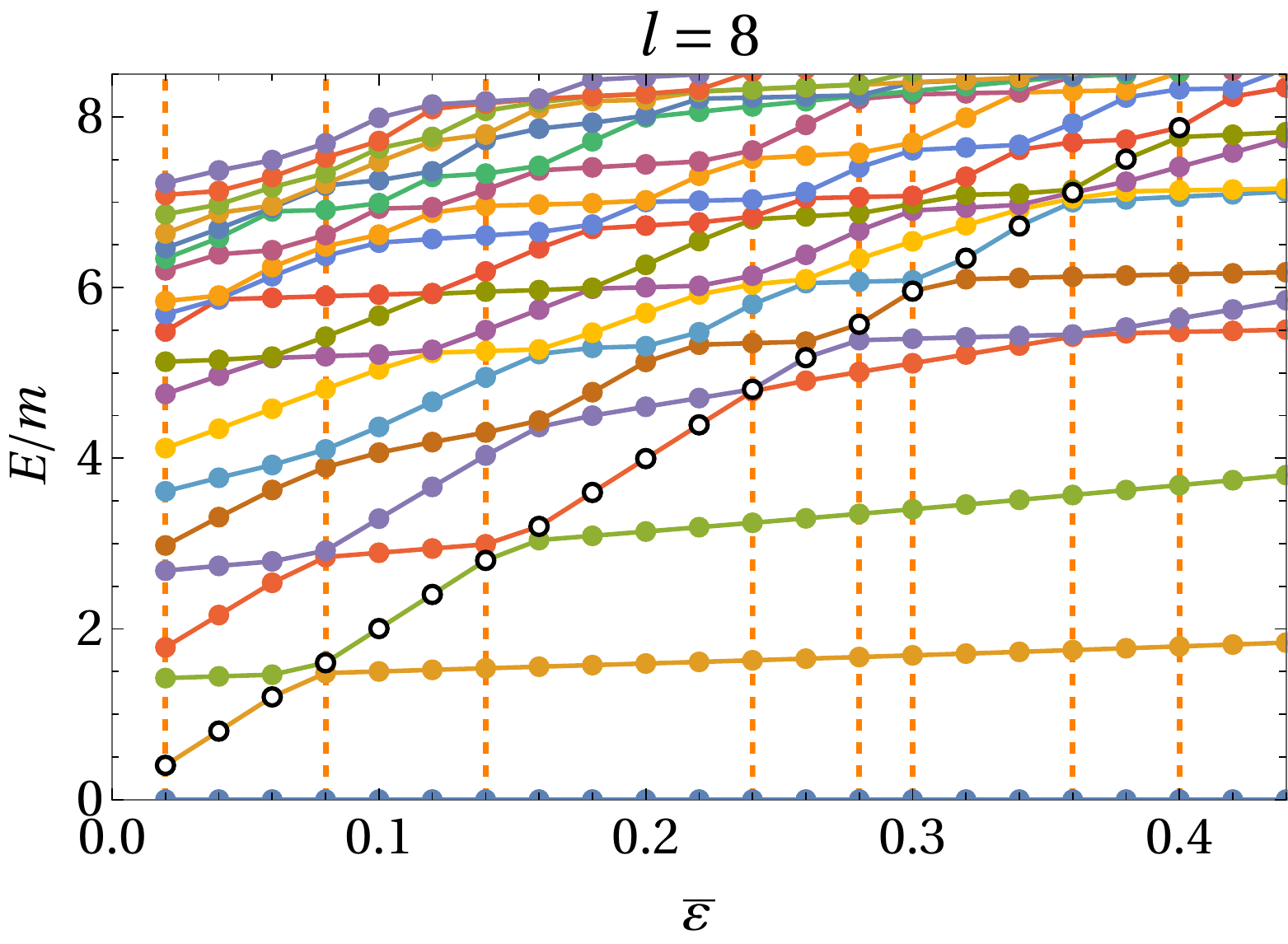}
    \caption{The spectrum of $H_\varepsilon$ for $l = 8$ and $\bar g = 1.1$ as a representative spectrum of the asymmetric theory, with the true vacuum line subtracted. The false vacuum line is denoted by open black markers.}
    \label{fig:asym-spectrum}
\end{figure}

\begin{figure}
    \centering
    \includegraphics[width = \linewidth]{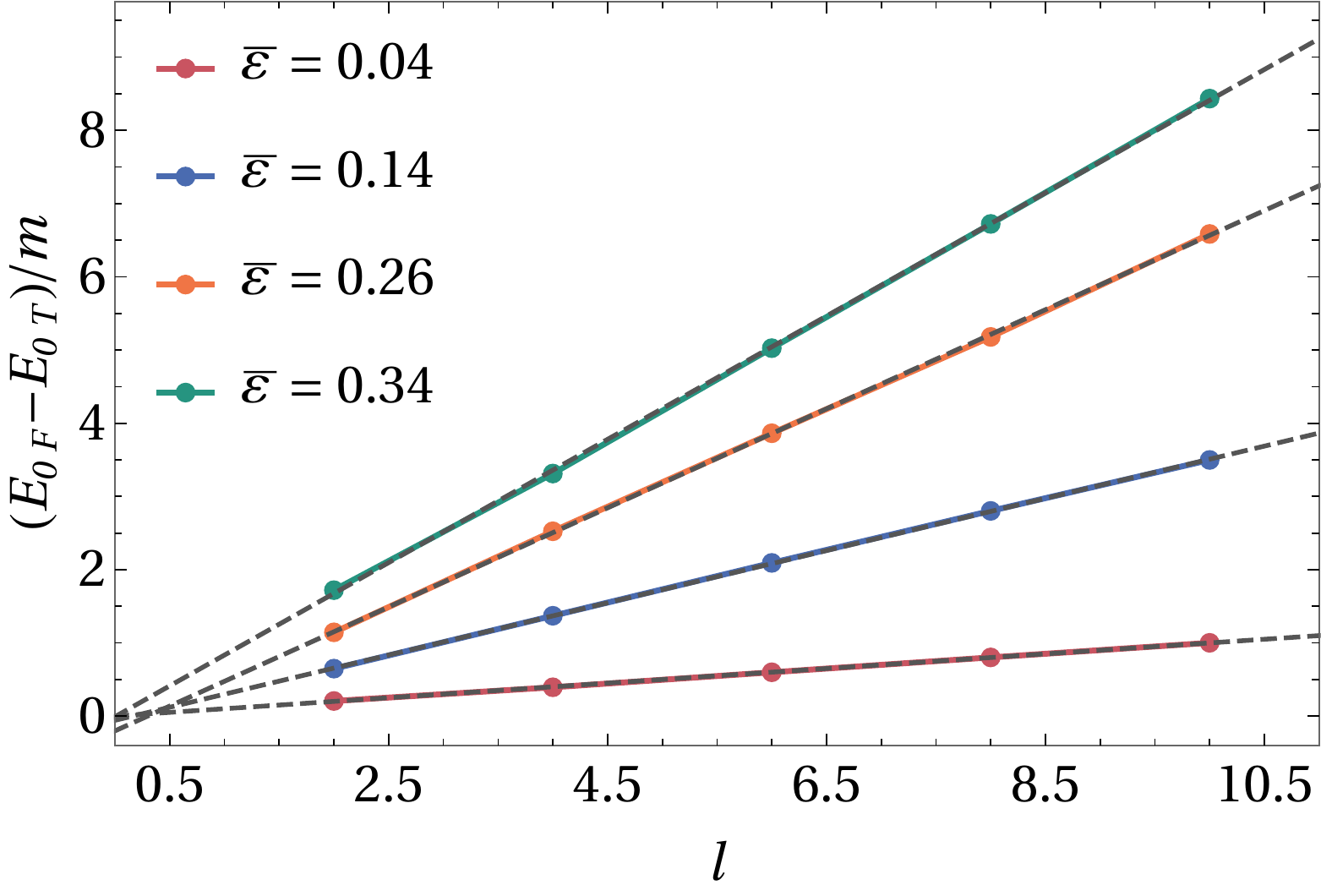}
    \caption{The energy difference between the false and true vacuum as a function of the volume with the linear dependence fitted for some values of the symmetry breaking field $\varepsilon$ for $\bar g = 1.1$. The slope of the fitted linear curve is the energy density difference (latent heat) $\mathcal{E}$ for a given $\bar\varepsilon$.}
    \label{fig:epsr-voldep}
\end{figure}

\begin{figure}
    \centering
    \includegraphics[width = \linewidth]{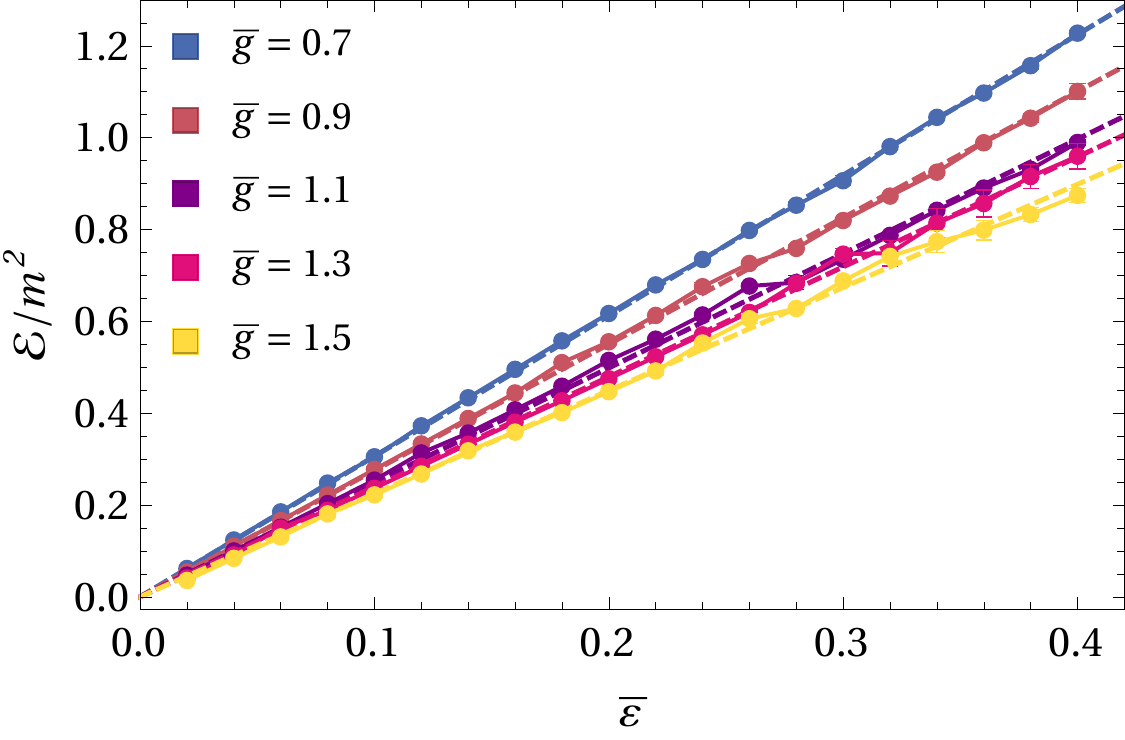}
    \caption{The latent heat $\mathcal{E}$ as a function of $\bar\varepsilon$ for various couplings $\bar g$ with the fitted linear dependence. The error bars represent a crude estimate obtained from the uncertainty of the parameter estimation from the fits shown in Fig. \ref{fig:epsr-voldep}.}
    \label{fig:epsr-fit}
\end{figure}

The latent heat $\mathcal{E}$ is the difference of energy density between the true and false vacuum, which can be easily computed from the THA diagonalising $H_\varepsilon$ by considering the spectrum as a function of $\bar\varepsilon$ for fixed volumes as illustrated in Fig. (\ref{fig:asym-spectrum}) for $\bar g=1.1$ and $l=8$. Performing this procedure for several values of the volume we can then plot the vacuum energy splitting as a function of the volume for each value of the symmetry breaking parameter $\varepsilon$, which is shown in \ref{fig:epsr-voldep} for $\bar g=1.1$. The slope of these lines gives the value of the latent heat $\mathcal{E}(\bar\varepsilon)$ as a function of $\bar\varepsilon$. For small $\varepsilon$ this function is expected to be linear corresponding to first order in perturbation theory, which turns out to hold in all the range of $\varepsilon$ needed for the later computations of the vacuum decay as shown in Fig. \ref{fig:epsr-fit} for various values of $\bar g$. As a result, the latent heat can be parameterised by fitting a linear relation 
\begin{equation}\label{eq:latent-heat}
    \mathcal{E} = A(\bar g)\bar\varepsilon
\end{equation}
and extracting the coefficient $A(\bar g)$ which only depends on $\bar g$. The procedure was carried out for several different values of $\bar g$, with a non-zero mode basis of around 1000 states and a minisuperspace dimension of 11, which provided sufficient accuracy as demonstrated by the results presented in the plots. 

\subsection{Bubble nucleation rate}
\begin{figure*}
    \centering
    \includegraphics[width = \linewidth]{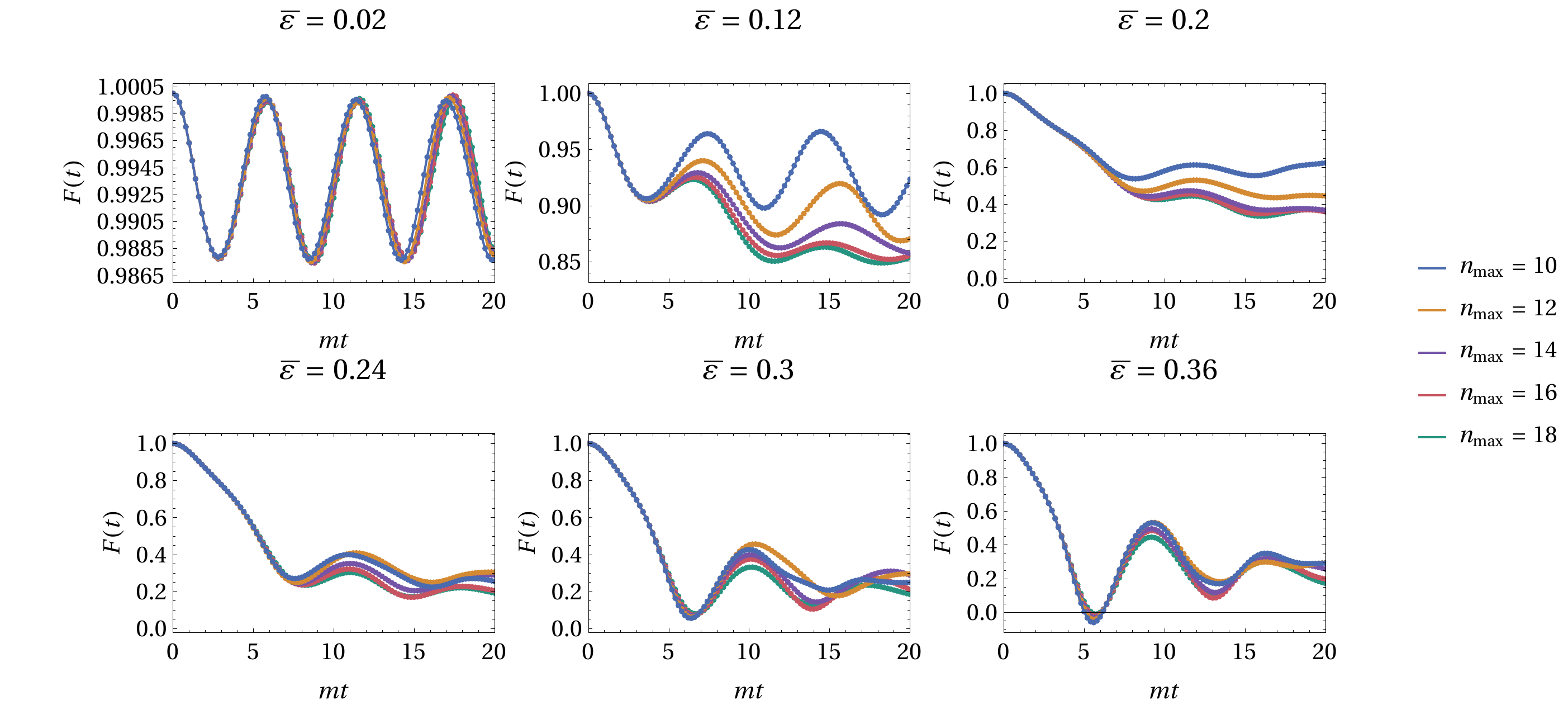}
    \caption{Time evolution of $F(t)$ for $\bar g = 1.1$ and different values of $\varepsilon$. The different colours denote different values of the cut-off. Here results corresponding to $n_{\text{max}} = 10, 12, 14, 16$ and $18$ are presented, corresponding to cc. $4100-287000$ states.}
    \label{fig:time-evo}
\end{figure*}

\begin{figure*}
    \centering
    \includegraphics[width = \linewidth]{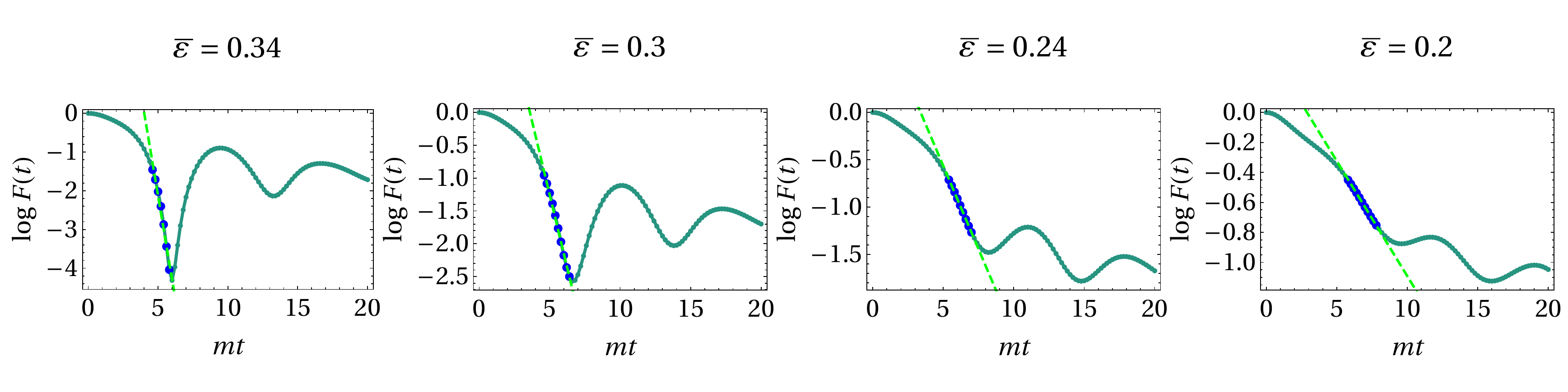}
    \caption{The logarithm of $F(t)$ for $l = 20$ and $\bar g = 1.1$ obtained with the largest cut-off $n_{\text{max}} = 18$, plotted together with the linear fits to the apparent tunnelling regime. The slope of the linear fits gives the tunnelling rate $\Gamma$.}
    \label{fig:log-fits}
\end{figure*}

Having computed the kink mass and the latent heat in the quantum theory, we turn to the evaluation of the tunnelling rate via the THA using the quantum quench setting presented in Section \ref{sec:2dphi4}. The tunnelling rate is computed as a function of the latent heat $\mathcal{E}$ which is controlled using the symmetry breaking parameter $\bar\varepsilon$. The validity of the THA simulation restricts the range of $\bar\varepsilon$ for which the simulation makes sense:
\begin{itemize}
    \item To avoid finite size effects, the size of the critical bubble \eqref{eq:resonant-bubble} must be smaller than the volume $L$, at least by a few times the correlation length (which for the regime of coupling considered here is of order $1/m$). 
    \item The false vacuum state must fit below the cut-off, therefore it is necessary to fulfil $\mathcal{E}=A(\bar g)\bar\varepsilon\ll \Lambda$.
\end{itemize} 
These two conditions impose a lower and upper limit on the values of $\bar\varepsilon$ for the simulations, which depend on the self-interaction $\bar g$. We simulated the time evolution in a volume $l = 20$, using a minisuperspace dimension of $41$ and verified that the results were stable against increasing the number of zero-mode eigenstates kept. For the non-zero modes, the cut-off was varied with the values $n_{\text{max}} = 10, 12, 14, 16$ and $18$, with the dimension of the truncated Hilbert space going up to cc.~$280000$. With these parameters and settings, it was possible to estimate the available interval for $\bar\varepsilon$ at any fixed choice for $\bar g$. Finally, we chose values for $\bar g$ for which this interval was large enough so that the dependence of the decay rate on $\bar\varepsilon$ could be seen in a reasonable interval.

Fig. \ref{fig:time-evo} shows the time evolution of $F(t)$ for coupling $\bar g = 1.1$ and different values of the symmetry breaking field $\bar\varepsilon$. For small values $\bar\varepsilon$ the size \eqref{eq:resonant-bubble} of the resonant bubble is too large compared to the volume, preventing nucleation and leading instead to persistent oscillations. For larger $\bar\varepsilon$ the nature of the time evolution changes: after a short initial transient corresponding to quantum Zeno regime\cite{Degasperis1974, Misra1977} where the time dependence is quadratic, a time window with exponential decay of $F(t)$ follows, which can be more easily identified as a linear drop on the plot of $\log F(t)$ shown in Fig. \ref{fig:log-fits}. The validity of exponential behaviour is limited in time, however, and it is followed by oscillations with their amplitude apparently decreasing in time with a power law, although the available time window is not long enough time to make this observation more precise. Note that even if the oscillating regime were absent, the range of time evolution available in the THA is limited from above by the volume, since for $t>L$ excitations can travel around the circle, resulting in deviations from infinite volume behaviour. Therefore the exponential behaviour can only be observed in a finite time window. We return to a more detailed discussion of the theoretical and methodological limitations of simulating the vacuum decay in the Conclusions.

\begin{figure}
    \centering
    \includegraphics[width = \linewidth]{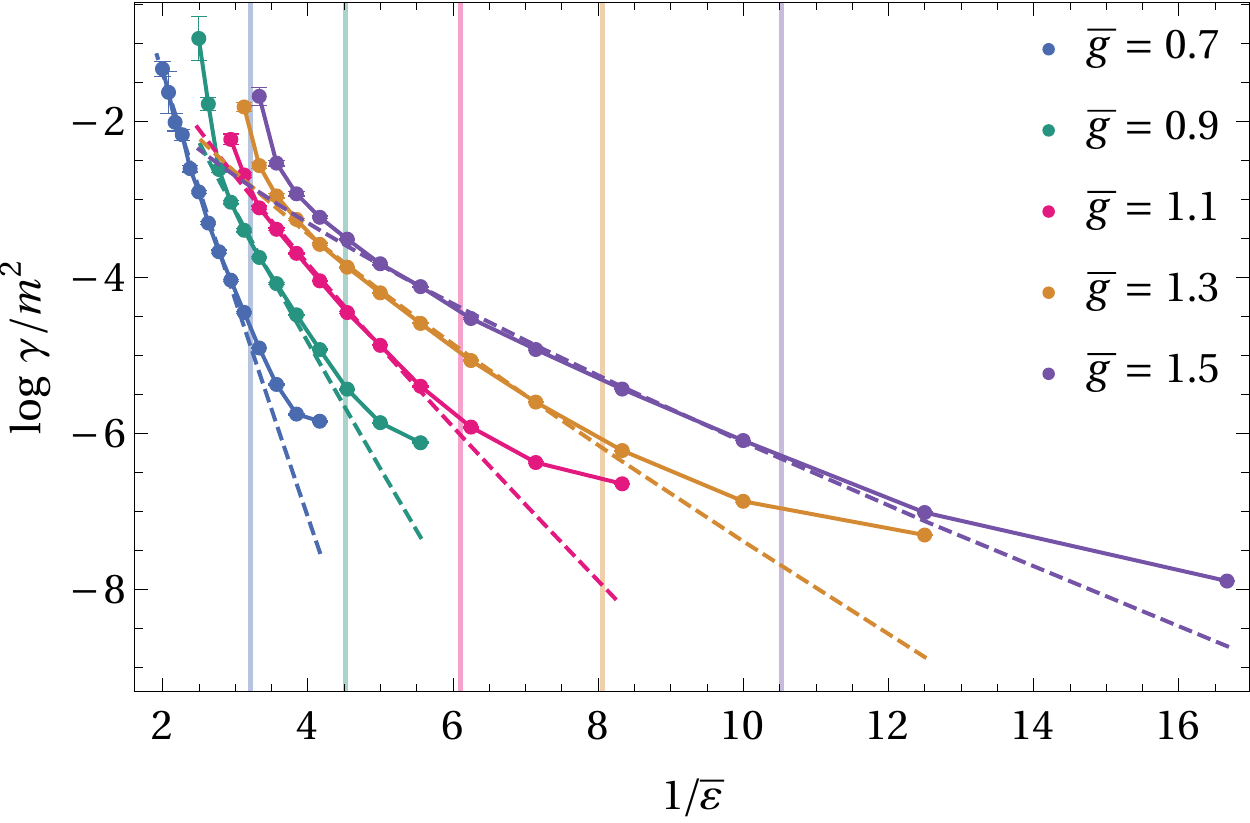}
    \caption{The logarithm of $\gamma$ obtained from THA (with $n_{\text{max}} = 18$) for various couplings in $l = 20$ as a function of $1/\bar\varepsilon$ together with the theoretical predictions (\ref{eq:voloshin}) (dashed lines). The vertical lines correspond to the values of the symmetry breaking field values where the resonant bubble size reaches $a_* = l/2$, demonstrating that the difference between the theoretical and numerical results for small values of $\bar\varepsilon$ originate from finite size effects. The error bars represent a crude estimate obtained from the uncertainty of the parameter estimation from the fits, a representative sample of which is shown in Fig. \ref{fig:log-fits}.}
    \label{fig:gamma-plot}
\end{figure}

To determine the tunnelling rate we plot the logarithm of $F(t)$ and determine the slope of the linear segment in the logarithmic plot. This was carried out for various quartic couplings $\bar g$ and $\bar\varepsilon$, with some representative plots shown in Fig. \ref{fig:log-fits} corresponding to $\bar g = 1.1$ and $l = 20$. Identification of the linear segment is simpler for larger values of $\bar\varepsilon$, while for smaller $\bar\varepsilon$ the identification is helped by following the time evolution gradually from larger to smaller values of the symmetry breaking field. Dividing by the value of $L$ gives the nucleation rate per unit volume which is shown in Fig. \ref{fig:gamma-plot} for $l = 20$ and various couplings as discrete data points connected by continuous lines for convenience. The dashed curves show the theoretical prediction computed using the kink mass $M$ and latent heat $\mathcal{E}$ extracted from THA using the formula
\begin{equation}\label{eq:voloshinA}
    \gamma = C(\bar g)\frac{\mathcal{E}}{2\pi} \exp{\left[-\frac{\pi M^2}{\mathcal{E}}\right]}\,,
\end{equation}
which differs from Voloshin's result \eqref{eq:voloshin} by including a $\bar g$-dependent factor $C(\bar g)$ which is a fitting parameter that can be used to translate the prediction curve to overlay it with the simulation results. Note that apart from this factor, the $\bar\varepsilon$ dependence follows the theoretical prediction very well. The appearance of such a redefinition is eventually expected since the same proved necessary when comparing simulation results for the transverse field Ising spin chain\cite{Lagnese2021} with the corresponding theoretical predictions \cite{Rutkevich1999}. 

The match between the numerical and theoretical results is made stronger by observing that it holds for different values of the coupling strength $\bar g$. In all cases, the curves show deviations both for small and for large values of $\bar\varepsilon$. For small values of $\bar\varepsilon$ (i.e. large values of $1/\bar\varepsilon$) the disagreement originates from finite size effects resulting from the size of the resonant bubble \eqref{eq:resonant-bubble} being comparable to the volume. The colour-coded vertical lines in Fig. \ref{fig:gamma-plot} are drawn at values of $\bar\varepsilon$ where the resonant bubble size $a_*$ is equal to half the volume. It can be clearly seen that this coincides well with the regime where the numerical results start to deviate appreciably from the theoretical predictions. 

The deviations for large values of the symmetry breaking field the numerical data can have two different origins. First, the theoretical predictions assume the thin-wall approximation which assumes a suitable small value of the symmetry breaking field, although we cannot really provide a concrete estimate for the value where the prediction should fail. In addition, for larger values of the latent heat the energy injected by the quantum quench becomes comparable with the truncation, leading to loss of precision of the numerical simulation. In addition, large values of the symmetry breaking field $\bar\varepsilon$ can even lead to the disappearance of the local minimum corresponding to the false vacuum, changing the dynamics entirely.

\begin{figure}
    \centering
    \includegraphics[width=\linewidth]{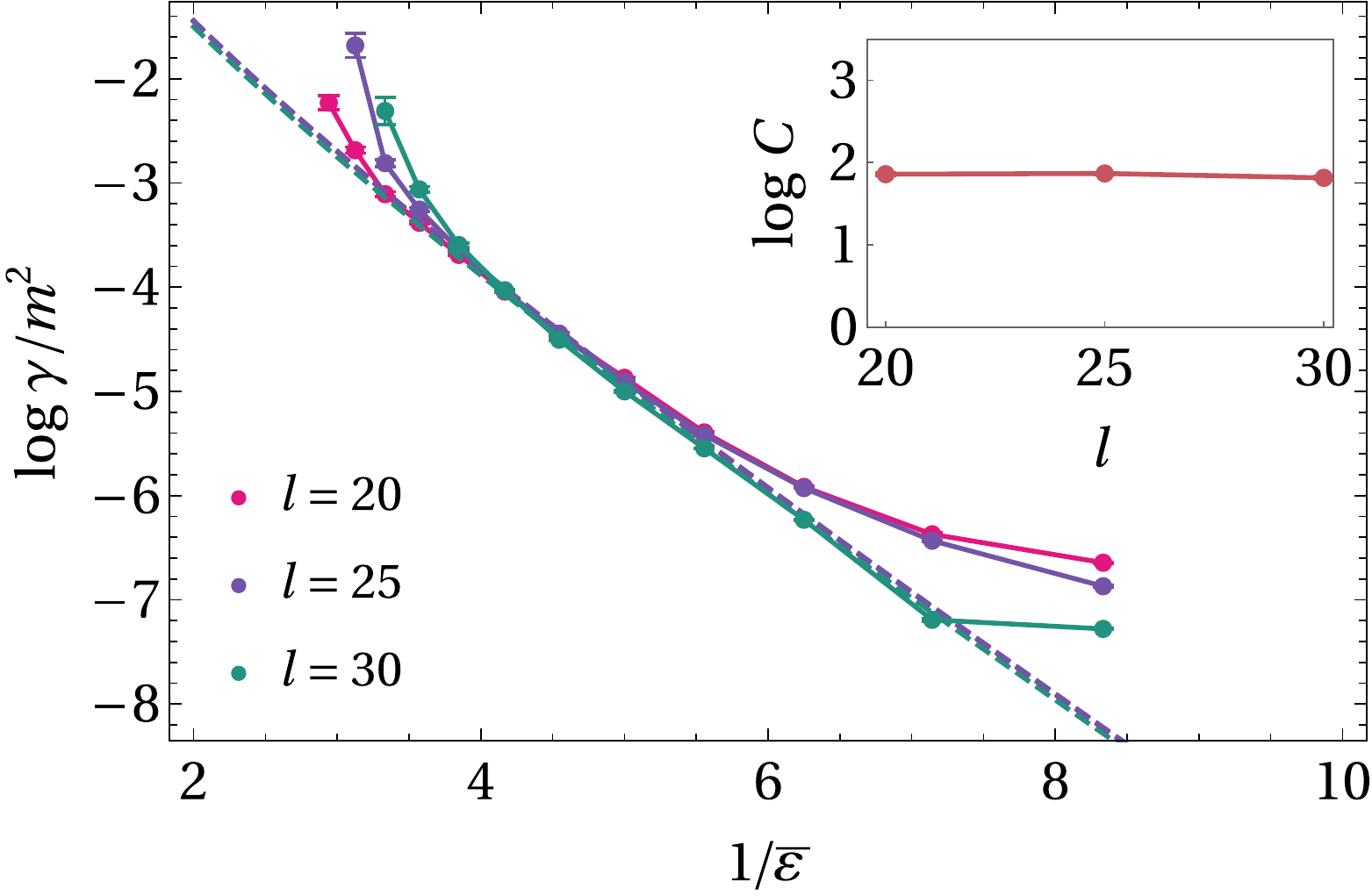}
    \caption{The (logarithm of the) nucleation rate $\gamma$ obtained from THA for different volumes $l$ at interaction $\bar{g} = 1.1$ as a function of $1/\bar\varepsilon$ together with the theoretical predictions (\ref{eq:voloshinA}). The inset shows the fitted values of $\log C$ for different values of the volume, where the (barely visible) error bars show the uncertainty of the parameter estimation from the fits.}
    \label{fig:gamma-plot-voldep}
\end{figure}

To test whether the factor $C(\bar{g})$ depends on the volume, we carried out simulations in other volumes  $l = 25$ and $30$. For these simulations we used a mini-superspace of dimension of $41$, together with a non-zero mode cutoff of $n_{\text{max}} = 18$, $22$ and $24$ resulting in Hilbert space dimensions of cc. $280000$, $680000$ and $680000$ for $l = 20$, $25$ and $30$, respectively. As illustrated in Fig. \ref{fig:gamma-plot-voldep}, the results show that $\log C(\bar{g})$ (which is the quantity directly obtained) shows no dependence on the volume. For large $1/\bar\varepsilon$, where the limiting factor is the finite system size, the agreement between the simulations and the prediction \eqref{eq:voloshinA} improves with larger volume, as expected. The deviations for small $1/\bar\varepsilon$ grow with the volume, which suggests that rather than resulting from the invalidity of the thin wall approximation, this discrepancy is caused by truncation effects, since according to \eqref{eq:ncutoff} the physical cut-off $\Lambda$ in units of $m$ is $4\pi n_{\text{max}}/l$, which in our simulations was decreasing with the volume.

\begin{figure}
    \centering
    \includegraphics[width=\linewidth]{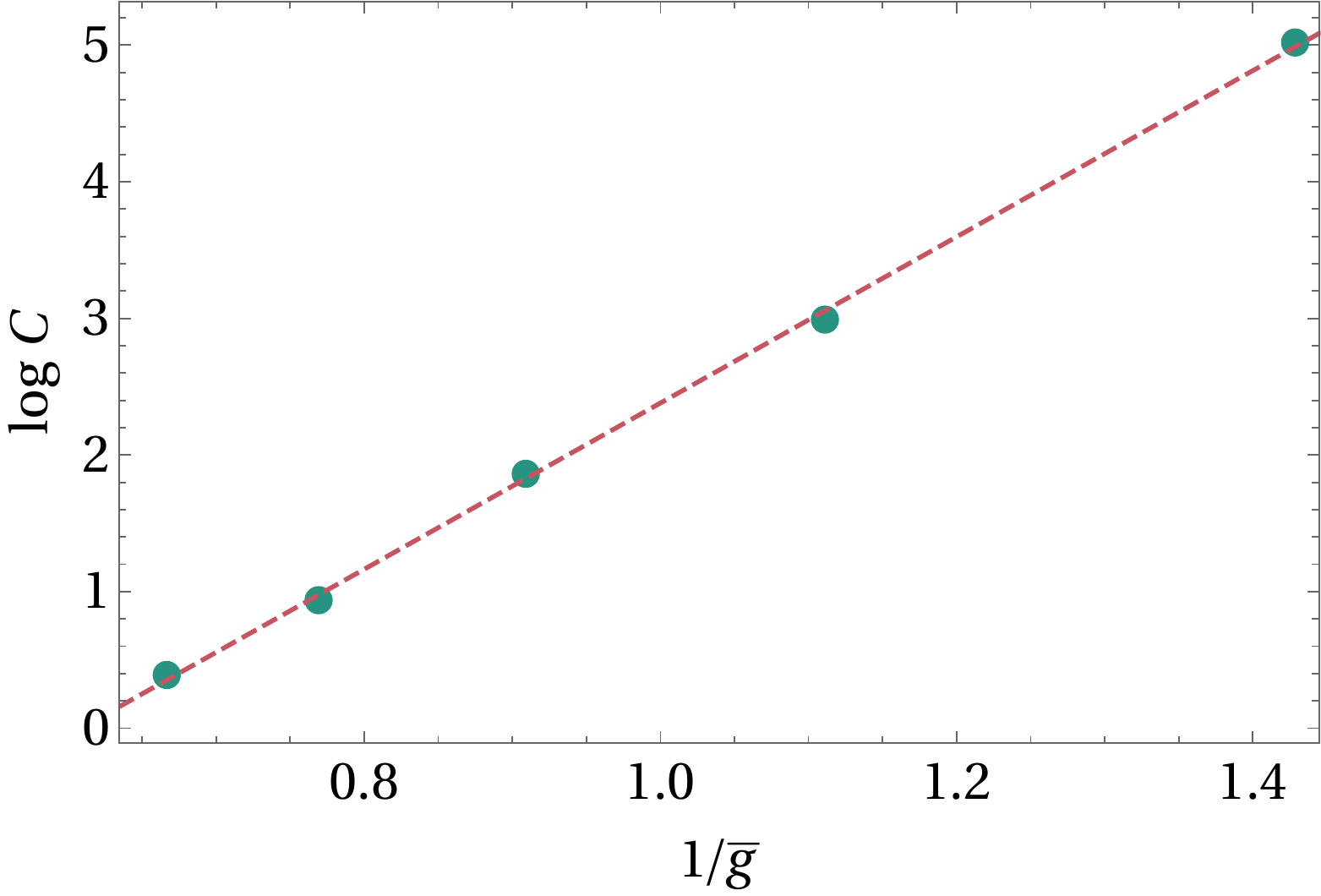}
    \caption{THA results (with $n_{\text{max}} = 18$) for the logarithm of the prefactor $C$ plotted as a function of the inverse of the interaction strength $1/\bar g$ together with a linear fit. The numerical evaluation was performed in volume $l = 20$.}
    \label{fig:logC-gdep}
\end{figure}

Fig. \ref{fig:logC-gdep} shows the numerically obtained values of $\log C$ as a function of  $1/\bar g$, extracted in volume $l=20$ as in Fig. \ref{fig:gamma-plot} which suggests that the correct prefactor in the nucleation rate has a highly non-perturbative dependence on the interaction strength. 

\section{Conclusions}\label{sec:conclusions}

We investigated the decay of the false vacuum in the 1+1 dimensional $\varphi^4$ theory by studying the time evolution of the order parameter triggered by a quantum quench. We simulated the dynamics by a truncated Hamiltonian approach built using the Fock space of the free massive boson as computational basis, and demonstrated the existence of a regime of exponential decay and extracted its rate. The decay rate normalised to unit volume was first computed theoretically in the semi-classical approximation by Coleman;\cite{Coleman1977, Coleman1977-2} here we used a later prediction by Voloshin\cite{Voloshin1985} which is expected to be exact at the quantum level if the nucleated bubbles are in the thin wall limit. Apart from an overall normalisation factor $C(\bar g)$, we found that the numerically determined decay rate considered as a function of the latent heat matches the theoretical predictions well. The appearance of $C(\bar g)$ is consistent with recent results obtained in the spin chain setting \cite{Lagnese2021}. Beyond the overall prefactor, the observed deviations are consistent with the expected limitations of the theoretical approach and the numerical simulation.

Extracting the nucleation rate from the decay rate of the order parameter is subject to certain general, as well as method specific limitations. Concerning the general limitations, at short times the exponential behaviour is absent due to general principles of quantum theory, while for later times a complicated dynamics takes place involving the expansion and collision of nucleated bubbles, and finally thermalisation of the resulting finite density medium \cite{Lagnese2021}. Specifically for the THA method our results demonstrate that despite the available time window is limited by the finite volume, it can still access the full time range in which the exponential behaviour holds.

As already mentioned in the Introduction, the decay of the false vacuum was recently studied in quantum spin chains using tensor network methods\cite{Lagnese2021}, for which time evolution can be simulated directly in infinite volume using tensor network methods, which is a definite advantage over the THA. However, in spite of the absence of finite size effects tensor network methods are still limited in their time range due to the buildup of entanglement. In addition, time evolution in spin chains is affected by lattice effects such as e.g. Bloch oscillations\cite{2020PhRvB.102d1118L} that can prevent the subsequent expansion, and therefore thermalisation, of the nucleated bubbles\cite{Pomponio2022}, while from the point of view of field theory the truncated Hamiltonian approach has the advantage that lattice effects are absent.

Now we turn to the issue of the appearance of the fitting parameter $C(\bar g)$. The predicted nucleation rate \eqref{eq:voloshin} has the form
\begin{equation}
    \gamma = \frac{\mathcal{E}}{2\pi} \exp{\left[-\frac{\pi M^2}{\mathcal{E}}\right]}\,,
\end{equation}
consisting of the exponential of the instanton action and a prefactor resulting from quantum fluctuations. The observed rate \eqref{eq:voloshinA} agrees with the above prediction if $C(\bar g)$ is set to $1$. In our comparison we eventually looked at the logarithm of the rate, whose dependence on the latent heat is dominated by the $1/\mathcal{E}$ coming from the instanton action; the presence and also the precise value of this contribution is strongly confirmed by the comparison to the simulation results. The fluctuations contribute a dependence $\log\mathcal{E}$ to $\log\gamma$, and while its presence is consistent with our simulation results, it cannot be verified to high precision due to the slowly changing nature of the logarithm. Nevertheless, the observed agreement strongly suggests that the relative factor $C(\bar g)$ between the theoretical predictions and numerical results is independent of the latent heat $\mathcal{E}$ as well as the volume $L$. The numerical data also indicate that $\log C(\bar g)$ is inversely proportional to the interaction strength $\bar{g}$. The precise origin of the factor $C(\bar g)$ is not clear to us at present and deserves further investigation. 

Other interesting avenues to explore is to extend our studies beyond the thin-wall regime, and also to other 1+1 dimensional quantum field theories. In addition, the late time behaviour eventually corresponds to thermalisation dynamics, which is another interesting physical regime to study in the future. 

\begin{acknowledgments}
This work was supported by the  National Research, Development and Innovation Office (NKFIH) through the OTKA Grants K 138606, and also within the Quantum Information National Laboratory of Hungary. 
\end{acknowledgments}

\bibliography{refs.bib}
\bibliographystyle{utphys.bst}

\end{document}